# A quasi-molecular mechanism of formation of hydrogen in the early Universe - a scheme of calculation

Tamaz Kereselidze[1] and Irakli Noselidze[2]

[1]Faculty of Exact and Natural Sciences, Tbilisi State University, Chavchavadze Avenue 3, 0179 Tbilisi, Georgia
[2]School of Science and Technology, University of Georgia, Kostava Str. 77a, 0171 Tbilisi, Georgia

E-mail: tamaz.kereselidze@tsu.ge

**Abstract**

In our recent papers (Kereslidze *et all* 2019a, 2021) a non-standard quasi-molecular mechanism was suggested and applied to treat the cosmological recombination. It was assumed that in the pre-recombination stage of evolution of the Universe an electron combined with two neighbouring protons and created the hydrogen molecular ion, $H_2^+$ in highly excited states, which then descended into the lower-lying states or dissociated.

In this work, we elaborate the scheme of calculation for free-bound radiative transitions into attractive states of $H_2^+$ as functions of redshift $z$. Together with the earlier developed treatment of bound-bound radiative transitions in $H_2^+$, the elaborated scheme of calculation can be used for the design of a fast and complete cosmological recombination code.

## 1. Introduction

Among various radiative and collisional processes occurring in natural and artificial plasmas, the terrestrial atmosphere and the interstellar medium the recombination of an electron and a proton plays an important role because this process was responsible for the formation of hydrogen atom in the early Universe. According to the pioneering works of Zel'dovich, Kurt and Syunyaev (1968) and Peebles (1968) charged electrons and protons first become bound to form electrically neutral hydrogen atoms in the recombination epoch of evolution of the Universe.

In cosmology the evolution of the Universe is characterised by redshift $z$, which is related to the temperature with equation $T = 2.725(1+z)$. As the Universe expanded, it concurrently cooled, eventually to a point at which the formation of neutral hydrogen atoms was favoured energetically. The recombination of hydrogen took place at redshift $800 \lesssim z \lesssim 2000$ when the energy of photons decreased below the ionization energy of hydrogen. When redshift decreased to $z \simeq 1100$ the temperature of matter and radiation decreased to $T \simeq 3000K$ and photons decoupled from matter in the Universe. After this time photons freely moved through the Universe, creating the cosmic microwave background radiation. For observations an important fact is that the cosmic microwave background spectrum experienced a unique distortion due to the release of photons during the recombination period of evolution of the Universe. These additional photons form together with the thermal spectrum a cosmological recombination spectrum. Despite substantial progress achieved in recent decades (Dubrovich and Grachev 2005, Chluba and Sunyaev 2006, Wong and Scott 2007, Grin and Hirata 2010, Ali-Haïmoud and Hirata 2010, Chluba and Thomas 2011, Chluba and Ali-Haïmoud 2016) there remain problems in understanding how the details of the recombination affect the cosmological parameters.

In our recent paper (Kereselidze *et al* 2021), to which we hereafter refer as paper I, a non-standard quasi-molecular mechanism of recombination (QMR) was suggested and applied to treat the cosmological recombination. According to the QMR, in the pre-recombination stage of evolution of the Universe, when the temperature and density of protons were higher than subsequently, the combination of an electron and a proton occurred in the presence of the nearest neighbouring proton, which participated in the process. In paper I (see also Kereselidze *et al* 2019a), we assumed that an electron collides with two protons situated one far from another, emits a photon and creates quasi-molecule $H_2^+$ in highly excited states. The main outcomes obtained in paper I were that the QMR allows the formation of $H_2^+$ in its ground state and that the probability of this process is comparable with the probability of formation of atomic hydrogen in the ground state. Furthermore, in paper I we showed that the QMR decreases the rate of recombination of hydrogen and shifts the beginning of the stage of the standard recombination toward an earlier period, i.e. larger redshift.

As an electron is much lighter than a proton, the velocity of electrons substantially exceeded a velocity of protons in the pre-recombination and recombination stages of evolution of the Universe. This fact allows us to treat the cosmological recombination on a basis of an adiabatic representation. In this approximation all characteristics of the process depend upon the distance $R$ between protons participating in the recombination. A quantitative analysis of the cosmological recombination requires a knowledge of wavefunctions that correctly describe an electron involved in the process in both the initial continuous and final discrete eigenstates. Several algorithms have been elaborated to calculate the discrete energy terms and wavefunctions of the quasi-molecule formed of an electron and two bare nuclei. For $H_2^+$, the most detailed description of an elaborated algorithm and the results of extensive calculations were presented by Bates and Reid (1968). This algorithm might serve to calculate wavefunctions describing an electron in the bound state. In our treatment we make use of wavefunctions of $H_2^+$ that are derived in an algebraic form at large distances $R$ between protons (Kereselidze *et al* 2003).

More challenging is to obtain the wavefunction that correctly describes an electron in the initial continuous spectral state. Wavefunctions of this type are mostly calculated numerically. Various computational methods, generally based on an infinite expansion of the wavefunctions in terms of some basis functions, have been developed to determine the desired eigenfunctions. An application of numerical wavefunctions to cosmological recombination involves cumbersome and tedious calculations and, accordingly, is time-consuming.

The variables in the Schrödinger equation with two fixed Coulomb centres are known to be separable in prolate spheroidal coordinates. In our paper (Kereselidze *et al* 2019b) the two-Coulomb-centre problem was solved for the continuous spectrum in prolate spheroidal coordinates; solutions of the one-dimensional equations obtained after separation of the variables were found in a closed algebraic form for large distances $R$ between the Coulomb centres. For an arbitrary configuration of protons the wavefunction of a colliding electron is representable as a linear combination of derived spheroidal functions. Despite representing spheroidal functions in a closed algebraic form, the application of their linear combinations to the cosmological recombination remains inconvenient. The problem hence requires an alternative treatment.

In the present work, we derive the two-Coulomb-centre wavefunction for the continuous spectrum in the form convenient for a quantitative analysis of the QMR. Our solution of the problem is based on the use of the nonrelativistic Coulomb Green's function (CGF) defined in parabolic coordinates. We apply the CGF to find the wavefunction of an electron that is moving in the field of one Coulomb centre and that experiences influence of another distant Coulomb centre. The derived wavefunction is applied for the calculation of probabilities of free-bound transitions as functions of redshift $z$. The obtained results will allow us to evaluate a contribution of the QMR in the formation of atomic hydrogen in the pre-recombination and recombination stages of evolution of the Universe and thereby to reveal an importance of the non-standard mechanism of recombination.

An advantage of the developed scheme of calculation is that in the sevenfold integral arising at the treatment the fivefold one can be calculated analytically. This fact is crucial for the elaboration of a fast and complete cosmological recombination code that includes the QMR. As the creation of $H_2^+$ in an excited repulsive state leads to the immediate dissociation and formation of atomic hydrogen (Kereselidze *et al* 2019a), which occurs analogously to the standard mechanism of recombination, we consider the transition of a colliding electron into an attractive state with a subsequent formation of $H_2^+$ in a long-lived vibrational state. The behaviour of the attractive energy terms of $H_2^+$ is depicted in Fig. 1; formulae describing these terms at large distances $R$ between protons are presented in paper I.

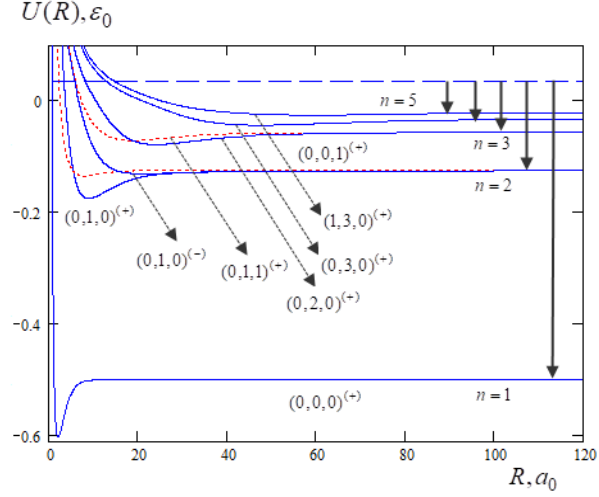

Figure 1. Some low-lying attractive energy terms of $H_2^+$ as functions of distance $R$ between protons. Energy terms are characterised by parabolic quantum numbers $(n_1, n_2, m_f)^{(\pm)}$ that specify electron states in the separate hydrogen atom; superscript $(\pm)$ denotes the symmetric and antisymmetric states of $H_2^+$ with respect to a reflection in the plane normal to and bisecting molecular axis. The solid and dotted curves correspond to $m_f = 0$ and $m_f = 1$, respectively. The dashed line denotes the energy of a colliding electron; $a_0 = \hbar^2 / m_e e^2 = 0.529 \times 10^{-8}$ cm is the first Bohr radius of hydrogen and $\varepsilon_0 = m_e e^4 / \hbar^2 = 27.21$ eV.

This paper is organized as follows. After stating our objective, we present the basic equations in section 2. The wavefunctions of an electron involved in the recombination are derived in section 3. The scheme of calculation, we present in section 4, before a conclusion in section 5. Unless otherwise indicated, atomic units ($e = m_e = \hbar = 1$) are used throughout the paper.

## 2. Basic equations

2.1 Average distance between protons

The average distance between protons at the beginning of the recombination period is estimable if one assumes that, before recombination, the reaction $e + p \rightleftarrows H + \hbar\omega$ was in statistical equilibrium, i.e. the rate of radiative recombination balanced the rate of photoionization. In statistical equilibrium at temperature $T$, the number density $n_a$ of particles with mass $m_a$ is given according to the Maxwell-Boltzmann equation. From this equation for the hydrogen atom, protons, and free electrons, we obtain the Saha-Boltzmann equation that relates the number densities of these particles:

$$\frac{n_p n_e}{n_H} = \left(\frac{m_e k_B T}{2\pi \hbar^2}\right)^{3/2} e^{-\frac{I}{k_B T}} . \qquad (1)$$

In (1) $k_B$ is the Boltzmann constant, $\hbar$ is the reduced Plank constant, and $I = 13.6$ eV is the ionization energy of the hydrogen atom.

The average distance between protons, $\bar{R}$ is related to the number density of protons according to $\bar{R} = n_p^{-1/3}$. Substituting in the latter equation $n_p$ defined from equation (1), in which the values of constants are inserted, we obtain $\bar{R}$ as a function of absolute temperature $T$ and ratio $n_H / n_e$

$$\bar{R}/cm = 7.45 \times 10^{-6} \left(\frac{n_H}{n_e}\right)^{-1/3} T^{-1/2} e^{\frac{52606}{T}}. \quad (2)$$

If we assume electrons to be in equilibrium with hydrogen atoms at the beginning of the recombination stage and, accordingly, to take $n_H = n_e$ in (2), we determine the dependence of $\bar{R}$ on redshift $z$. The deviation of this obtained dependence from that calculated with recombination code COSMOREC (Chluba and Thomas 2011) is important for $z \lesssim 1500$. For a larger redshift, equation (2) hence reproduces satisfactorily the dependence of $\bar{R}$ on $z$. Making the appropriate calculations, we thus obtain that for $z \gtrsim 1500$ the average distance between protons is comparable with radius $r_n$ of the hydrogen atom in a highly excited state, $\bar{R} \simeq r_n \gg 1$ (see fig. 1 in Kereselidze *et al* 2019a). The influence of a nearest neighbouring proton on the electron-proton recombination must hence be significant at redshift $z \gtrsim 1500$ and, accordingly, should be taken into account.

2.2 The Coulomb Green's function

The application of the CGF to investigate various radiative and collisional processes is not new. Since the beginning of 1960s, many papers on the properties and applications of the CGF have been published (Hostler 1962, 1964, Hostler and Pratt 1963, Kereselidze and Chibisov 1975, Blinder 1981, Chetouani and Hamman 1987, Swainson and Drake 1991, Maquet 1998, Laha 2005, Zaytsev *et al* 2020).

The CGF can be constructed from its spectral representation

$$G^{(+)}(\vec{r}, \vec{r}') = \sum_n \frac{\psi_n^*(\vec{r})\psi_n(\vec{r}')}{E_n - E}, \quad (3)$$

in which the summation runs over the complete set of discrete and continuum eigenstates; symbol (+) corresponds to an outgoing wave when $\vec{r} \to \infty$. In the first attempt to evaluate the CGF, Meixner (1933) tried to evaluate this function by explicit summation over eigenfunctions in parabolic coordinates. Further progress was made possible with an integral representation for a product of two Whittaker functions. Hostler in 1962 obtained the general closed-form expression for $G^{(+)}(\vec{r}, \vec{r}')$ on summing over Coulomb eigenfunctions in spherical polar coordinates. Blinder (1981) showed that summation (3) explicitly written in terms of discrete and continuous eigenstates in parabolic coordinates leads to the integral representation of the CGF.

Making use of the scheme of calculation developed by Blinder, we evaluate the CGF in the form convenient for our calculations:

$$G^{(+)}(\vec{r}, \vec{r}') = -\frac{ik}{2\pi} \sum_{m=-\infty}^{\infty} e^{im(\varphi-\varphi')} \int_0^\infty ds \, e^{i\frac{k}{2}(\mu+\nu+\mu'+\nu')\cosh s}$$
$$\cdot \sinh s \left(\coth \frac{s}{2}\right)^{2iz_1/k} J_m\left(k(\mu\mu')^{1/2} \sinh s\right) J_m\left(-k(\nu\nu')^{1/2} \sinh s\right). \quad (4)$$

In (4) $\mu = r(1+\cos\vartheta)$, $\nu = r(1-\cos\vartheta)$, $\varphi = \text{arctg}(y/x)$ are parabolic coordinates, in which $r$ is the radial variable and $\vartheta$ is the polar angle. Details of the derivation of equation (4) are presented in appendix A.

**3. Wavefunctions of an electron**

To begin, we find the wavefunction of an electron that collides with two fixed protons $a$ and $b$. At large distances $R$ between protons the wavefunction of an electron is representable as

$$\Psi_i^{(\pm)} = \frac{1}{\sqrt{2}}\left(\psi^{(a)} \pm \psi^{(b)}\right). \tag{5}$$

In (5) $\psi^{(a)}$ ($\psi^{(b)}$) is the wavefunction of an electron moving in the Coulomb field of proton $a$ ($b$) and that is perturbed by another proton. For the definiteness, we derive wavefunction $\psi^{(a)}$ that is centred on proton $a$. The appropriate Schrödinger equation reads

$$\left(-\frac{1}{2}\Delta_{\vec{r}} - \frac{1}{r} - \frac{k_0^2}{2}\right)\psi^{(a)}(r) = \frac{1}{|\vec{R}-\vec{r}|}\psi^{(a)}(\vec{r}). \tag{6}$$

Here, $\vec{r}$ is the position-vector of an electron with respect to proton $a$, $k_0^2/2$ is the electron energy and $R \gg 1$ is the distance between protons; $\vec{k}_0$ is the wave-vector directed along axis $z$, vector $\vec{R}$ is directed from proton $a$ to proton $b$.

Our purpose is to find the solution of equation (6) for $r < R$; thereby we exclude from the consideration the region near proton $b$. As a result term $|\vec{R}-\vec{r}|^{-1}$ in the right side of equation (6) can be expanded in powers of $r/R$. Considering only the first two terms of this expansion equation (6) takes the form

$$\left(-\frac{1}{2}\Delta_{\vec{r}} - \frac{1}{r} - \frac{k^2}{2}\right)\psi(r) = \upsilon(\vec{r})\psi(\vec{r}), \tag{7}$$

in which $k^2 = k_0^2 + 2/R$, $\upsilon = \vec{n}\vec{r}/R^2$ and $\vec{n} = \vec{R}/R$. In (7) and in following equations superscript $a$ is omitted in wavefunction $\psi^{(a)}$.

Introducing the CGF as the solution of the inhomogeneous different equation

$$\left(-\frac{1}{2}\Delta_{\vec{r}} - \frac{1}{r} - \frac{k^2}{2}\right)G^{(+)}(\vec{r},\vec{r}') = \delta(\vec{r}-\vec{r}'), \tag{8}$$

the eigenfunction of equation (7) that satisfies the appropriate boundary conditions is expressible as

$$\psi_{\vec{k}}(\vec{r}) = \psi_{\vec{k}}^{(0)}(\vec{r}) + \int G^{(+)}(\vec{r},\vec{r}')\upsilon(r')\psi_{\vec{k}}(\vec{r}')d\vec{r}'. \tag{9}$$

Here, $\psi_{\vec{k}}^{(0)}(\vec{r})$ is the solution of equation (7) with the zero right side. The solution of the homogeneous equation that is normalized with the delta function reads (Landau and Lifshitz 1977)

$$\psi_{\vec{k}}^{(0)}(\vec{r}) = N_i e^{i\vec{k}\vec{r}} {}_1F_1\left(i/k, 1, i(kr-\vec{k}\vec{r})\right),$$
$$N_i = (2\pi)^{-3/2} e^{\pi/2k} \Gamma(1-i/k), \tag{10}$$

in which ${}_1F_1\left(iZ_1/k, 1, i(kr-\vec{k}\vec{r})\right)$ is a confluent hypergeometric function, $N_i$ is a normalising factor and $\Gamma(1-i/k)$ is a gamma function.

In equation (9) function $\upsilon(\vec{r}')$ is of order of $R^{-2}$ in the region near nucleus $a$, increases when $\vec{r}'$ increases, and becomes of order of $R^{-1}$ near of nucleus $b$. On the contrary, function $\psi(\vec{r}')$ is oscillatory with decreasing amplitude as variable $\vec{r}'$ increases (Kereselidze et al 2019b). Hence in the right side of equation (9) the second term is much smaller than the first one and, accordingly, can be considered as perturbation. Replacing $\psi_{\vec{k}}(\vec{r}')$ by $\psi_{\vec{k}}^{(0)}(\vec{r}')$ in (9), we thereby obtain the desired wavefunction in the Coulomb-Born approximation

$$\psi_{\vec{k}}(\vec{r}) = \psi_{\vec{k}}^{(0)}(\vec{r}) + \frac{N_i}{R^2}\int G^{(+)}(\vec{r},\vec{r}')(\vec{n}\vec{r}')e^{i\vec{k}\vec{r}'} F\left(\frac{i}{k}, 1, i(kr'-\vec{k}\vec{r}')\right)d\vec{r}'. \tag{11}$$

In (11) the first unperturbed term is of order of one, whereas the second perturbed term is $O(R^{-1})$.

The wavefunction centred on proton $b$ can be found in an analogous manner.

In parabolic coordinates

$$\vec{n}\vec{r}' = \frac{\mu' - \nu'}{2}\cos\vartheta_{\vec{R}} + \sqrt{\mu'\nu'}\cos\varphi'\sin\vartheta_{\vec{R}}, \tag{12}$$

where $\vartheta_{\vec{R}}$ is the angle between $\vec{R}$ and $\vec{k}$; azimuthal angles $\varphi'$ and $\varphi_{\vec{R}}$ are measured from plane $(\vec{k},\vec{R})$, accordingly, $\varphi_{\vec{R}} = 0$ in (12). Inserting (4) and (10) in (11) together with (12) and performing the integration over $\varphi'$, we obtain the wavefunction of a colliding electron in parabolic coordinates

$$\psi_{\vec{k}} = \psi_{\vec{k}}^{(0)}(\mu,\nu) - \frac{ikN_i}{8R}\Big[f_1(\mu,\nu)\cos\vartheta_{\vec{R}} + f_2(\mu,\nu)\cos\varphi\sin\vartheta_{\vec{R}}\Big], \tag{13}$$

in which

$$f_1 = R^{-1}\int_0^\infty ds\,\sinh s\left(\coth\frac{s}{2}\right)^{2i/k}\int_0^{\mu'_{\max}}\int_0^{\nu'_{\max}} e^{i\frac{k}{2}\left[(\mu+\nu+\mu'+\nu')\cosh s + \mu' - \nu'\right]}$$

$$\cdot J_0\Big(k(\mu\mu')^{1/2}\sinh s\Big)J_0\Big(-k(\nu\nu')^{1/2}\sinh s\Big)F(i/k,1,ik\nu')(\mu'^2 - \nu'^2)d\mu'd\nu', \tag{14}$$

$$f_2 = 2R^{-1}\int_0^\infty ds\,\sinh s\left(\coth\frac{s}{2}\right)^{2i/k}\int_0^{\mu'_{\max}}\int_0^{\nu'_{\max}} e^{i\frac{k}{2}\left[(\mu+\nu+\mu'+\nu')\cosh s + \mu' - \nu'\right]}$$

$$\cdot J_1\Big(k(\mu\mu')^{1/2}\sinh s\Big)J_1\Big(-k(\nu\nu')^{1/2}\sinh s\Big)F(i/k,1,ik\nu')\sqrt{\mu'\nu'}(\mu' + \nu')d\mu'd\nu'.$$

In (14) the upper limits of integration are chosen from the condition $r'_{\max} = (\mu'_{\max} + \nu'_{\max})/2 \lesssim R/2$.

We proceed to find the wavefunctions of a bound electron. For this purpose we introduce a rotating coordinate system $(\tilde{x},\tilde{y},\tilde{z})$. If we assume that axis $\tilde{z}$ is directed along $\vec{R}$ and that protons are located on this axis with coordinates $\tilde{z}_a = 0$ and $\tilde{z}_b = R$, the wavefunction of a bound electron that corresponds to an attractive state can be written as a sum or difference of the appropriate wavefunctions centred on each nucleus (Bates and Reid 1968)

$$\Psi_f^{(+)} = \frac{1}{\sqrt{2}}\Big(\psi_{n_1,n_2,0}^{(a)}(\tilde{\mu}_a,\tilde{\nu}_a,\tilde{\varphi}) + \psi_{n_1,n_2,0}^{(b)}(\tilde{\mu}_b,\tilde{\nu}_b,\tilde{\varphi})\Big),$$

$$\Psi_f^{(-)} = \frac{1}{\sqrt{2}}\Big(\psi_{n_1,n_2,\pm 1}^{(a)}(\tilde{\mu}_a,\tilde{\nu}_a,\tilde{\varphi}) - \psi_{n_1,n_2,\pm 1}^{(b)}(\tilde{\mu}_b,\tilde{\nu}_b,\tilde{\varphi})\Big). \tag{15}$$

In (15) $R \gg 2n^2$, $n_1$ and $n_2$ are parabolic quantum numbers that specify electron states in the isolated hydrogen atom, $m_f = 0,1$ denotes the absolute value of the projected orbital angular momentum of an electron along the axis $\tilde{z}$ of the rotating coordinate system, and $n = n_1 + n_2 + m_f + 1$.

When the internuclear distance in $H_2^+$ is greater than the size of the shell on either nucleus, the wavefunction centred on nucleus $a$ is representable in the rotating coordinate system as (Kereselidze et al 2003)

$$\psi_{n_1,n_2,\pm m_f} = N_f X_{n_1,m_f}(\tilde{\mu})Y_{n_2,m_f}(\tilde{\nu})e^{\pm im_f\tilde{\varphi}}/\sqrt{2\pi}, \tag{16}$$

in which

$$X_{n_1,m_f} = e^{-\frac{\gamma}{2}\tilde{\mu}}\tilde{\mu}^{\frac{m_f}{2}}F\left(-n_1,m_f+1,\gamma\alpha_1\tilde{\mu}\right)\left(1 + \frac{(2n+2n_2+m_f)\tilde{\mu}}{4R}\right) + O(R^{-2}),$$

$$Y_{n_2,m_f} = e^{-\frac{\gamma}{2}\tilde{\nu}}\tilde{\nu}^{\frac{m_f}{2}}F\left(-n_2,m_f+1,\gamma\alpha_2\tilde{\nu}\right)\left(1 + \frac{(2n-2n_1-m_f)\tilde{\nu}}{4R}\right) + O(R^{-2}), \tag{17}$$

$\gamma = \sqrt{-2\varepsilon(R)}$, $\varepsilon(R) < 0$ is the energy of an electron in $H_2^+$, $\alpha_{1,2} = 1 \mp n(2n_{2,1} + m_f + 1 \pm 2n)/2R$ and $N_f$ is a normalising factor. The presented wavefunction is valid in the main region of the distribution of a bound electron, i.e. in the region in which $\tilde{\mu},\tilde{\nu} < R/2$.

We rewrite wavefunction (17) in variables $\mu, \nu, \varphi$ that are defined in the fixed coordinate system. Parabolic coordinates in the rotating and fixed coordinate systems are related by relations

$$\tilde{\mu} = \frac{\mu+\nu}{2} + \frac{\mu-\nu}{2}\cos\vartheta_{\vec{R}} + \sqrt{\mu\nu}\cos\varphi\sin\vartheta_{\vec{R}},$$

$$\tilde{\nu} = \frac{\mu+\nu}{2} - \frac{\mu-\nu}{2}\cos\vartheta_{\vec{R}} - \sqrt{\mu\nu}\cos\varphi\sin\vartheta_{\vec{R}}, \quad (18)$$

$$\tilde{\varphi} = \arctan\frac{\sqrt{\mu\nu}\sin\varphi}{\sqrt{\mu\nu}\cos\varphi\cos\vartheta_{\vec{R}} - \frac{\mu-\nu}{2}\sin\vartheta_{\vec{R}}}.$$

In variables $\mu, \nu, \varphi$ wavefunction (16) reads

$$\psi_{n_1,n_2,\pm m_f} = N_f e^{-\frac{\gamma}{2}(\mu+\nu)} u^{\frac{m_f}{2}} F(-n_1, m_f+1, \gamma\alpha_1\tilde{\mu})$$
$$\cdot F(-n_2, m_f+1, \gamma\alpha_2\tilde{\nu})\left[1 + \frac{1}{2R}Q(\tilde{\mu},\tilde{\nu})\right]e^{\pm im_f\tilde{\varphi}}/\sqrt{2\pi}, \quad (19)$$

in which

$$u = \left(\frac{\mu+\nu}{2}\right)^2 - \left(\frac{\mu-\nu}{2}\cos\vartheta_{\vec{R}} + \sqrt{\mu\nu}\cos\varphi\sin\vartheta_{\vec{R}}\right)^2,$$

$$Q = n(\tilde{\mu}+\tilde{\nu}) + n_2\tilde{\mu} - n_1\tilde{\nu} + \frac{m_f}{2}(\tilde{\mu}-\tilde{\nu}), \quad (20)$$

and $\tilde{\mu}, \tilde{\nu}, \tilde{\varphi}$ are defined with equations (18). We note that, in the wavefunction of a bound electron, the dependence on an orientation of protons arises already in the unperturbed term, whereas this dependence appears in the perturbed term in the wavefunction of a colliding electron.

## 4. Scheme of calculation

In the fixed coordinate system the operator of electric-dipole strength is $\vec{d} = -(\vec{i}x + \vec{j}y + \vec{k}z)$. For convenience, we calculate the matrix elements of operators $d_{\pm} = -(x \pm iy)$ and $d_z = -z$. In parabolic coordinates these operators read $d_{\pm} = -\sqrt{\mu\nu}e^{\pm i\varphi}$ and $d_z = -(\mu-\nu)/2$.

Matrix elements $(d_z)_{if}$ and $(d_{\pm})_{if}$ calculated over wavefunctions (13) and (19) are representable as

$$(d_z)_{i,f} = -\frac{N_f}{8}\left[U_{n_1,n_2,\pm m_f}(\vartheta_{\vec{R}}) + \frac{1}{2R}V_{n_1,n_2,\pm m_f}(\vartheta_{\vec{R}})\right],$$

$$(d_{\pm})_{i,f} = -\frac{N_f}{4}\left[U^{(\pm)}_{n_1,n_2,\pm m_f}(\vartheta_{\vec{R}}) + \frac{1}{2R}V^{(\pm)}_{n_1,n_2,\pm m_f}(\vartheta_{\vec{R}})\right], \quad (21)$$

in which

$$U_{n_1,n_2,\pm m_f} = \int_0^\infty\int_0^\infty e^{-\frac{\gamma}{2}(\mu+\nu)}\psi^{(0)}_{\vec{k}}(\mu,\nu)A_{n_1,n_2,\pm m_f}(\mu,\nu)(\mu^2-\nu^2)d\mu d\nu,$$

$$V_{n_1,n_2,\pm m_f} = \int_0^\infty\int_0^\infty e^{-\frac{\gamma}{2}(\mu+\nu)}\Big\{\psi^{(0)}_{\vec{k}}(\mu,\nu)B_{n_1,n_2,\pm m_f}(\mu,\nu) - \frac{ikN_i}{4R}[f_1(\mu,\nu)$$
$$\cdot A_{n_1,n_2,\pm m_f}(\mu,\nu)\cos\vartheta_{\vec{R}} + f_2(\mu,\nu)C_{n_1,n_2,\pm m}(\mu,\nu)\sin\vartheta_{\vec{R}}]\Big\}(\mu^2-\nu^2)d\mu d\nu,$$

and $\quad (22)$

$$U^{(\pm)}_{n_1,n_2,\pm m_f} = \int_0^\infty\int_0^\infty e^{-\frac{\gamma}{2}(\mu+\nu)}\psi^{(0)}_{\vec{k}}(\mu,\nu)A^{(\pm)}_{n_1,n_2,\pm m_f}(\mu,\nu)\sqrt{\mu\nu}(\mu+\nu)d\mu d\nu,$$

$$V^{(\pm)}_{n_1,n_2,\pm m_f} = \int_0^\infty \int_0^\infty e^{-\frac{\gamma}{2}(\mu+\nu)} \left\{ \psi^{(0)}_{\vec{k}}(\mu,\nu) B^{(\pm)}_{n_1,n_2,\pm m_f}(\mu,\nu) - \frac{ikN_i}{4R}\left[ f_1(\mu,\nu) \right. \right.$$
$$\left. \left. \cdot A^{(\pm)}_{n_1,n_2,\pm m_f}(\mu,\nu) \cos\vartheta_{\vec{R}} + f_2(\mu,\nu) C^{(\pm)}_{n_1,n_2,\pm m_f} \sin\vartheta_{\vec{R}} \right] \right\} \sqrt{\mu\nu}(\mu+\nu) d\mu d\nu.$$

Because the wavefunction of a bound electron rapidly decreases when $\mu$ and $\nu$ increase, the integration is extended up to infinity in (22). Functions $A_{n_1,n_2,\pm m_f}$, $B_{n_1,n_2,\pm m_f}$, $C_{n_1,n_2,\pm m_f}$ and $A^{(\pm)}_{n_1,n_2,\pm m_f}$, $B^{(\pm)}_{n_1,n_2,\pm m_f}$, $C^{(\pm)}_{n_1,n_2,\pm m_f}$ are defined in appendix B.

The probability of a free-bound radiative transition depends on the distance $R$ between protons and is defined as (Heitler 1954)

$$W_{i,f}(R) = \frac{4\omega^3_{i,f}(R)}{3c^3} \left|\vec{d}_{i,f}(R)\right|^2. \tag{23}$$

Here $\omega_{if}$ is the frequency of an emitted photon, $c$ is speed of light, and $\vec{d}_{if}$ are transition matrix elements. Taking into account that

$$\left|\vec{d}_{i,f}\right|^2 = \frac{1}{2}\left( \left|(d_+)_{i,f}\right|^2 + \left|(d_-)_{i,f}\right|^2 + 2\left|(d_z)_{i,f}\right|^2 \right), \tag{24}$$

and assuming that $R \equiv \bar{R}$ in (23), with $\bar{R}$ as the average distance between protons that is defined by equation (2), we thereby determine the free-bound transition probability as a function of redshift $z$.

Our purpose is to calculate transition probabilities at the parallel ($\vartheta_{\vec{R}} = 0$) and perpendicular ($\vartheta_{\vec{R}} = \pi/2$) orientation of protons with respect to the direction of propagation of a colliding electron. In this case, as shown in appendixes $B$ and $C$, the integration over $\varphi$, $\varphi'$ and $\mu$, $\mu'$, $\nu$ can be carried out analytically. The remaining integration over $\nu'$ and $s$ is to be performed numerically.

## 5. Conclusion

The cosmological recombination radiation constitutes a fundamental signal from the early Universe. Its detailed study can provide an efficient way to obtain some of the key cosmological parameters and allow testing physical processes that took place in the early Universe. Nowadays, the recombination history can be computed using the fast and detailed computation code COSMOSPEC (Chluba and Ali-Haïmoud 2016). COSMOSPEC includes important radiative transfer effects, 500-shell free-bound and bound-bound emission for hydrogen, and is based on the earlier designed software package COSMOREC (Chluba and Thomas 2011). It thus becomes clear that for a comprehensive study of the cosmological recombination all important physical processes must be incorporated in treatment.

While in terms of the standard theory (Zel'dovich, Kurt and Syunyaev 1968, Peebles 1968), the thermal history of the Universe is expected to be well understood, non-standard processes can modify this history. In the pre-recombination stage of evolution of the Universe, when the temperature and density of protons were higher than subsequently, an electron combined with two protons and created quasi-molecule $H_2^+$ in highly excited states. This non-standard quasi-molecular mechanism of recombination decreases the rate of recombination of hydrogen and shifts the beginning of the stage of the standard recombination toward an earlier period. It is thus extremely important to understand how the cosmological recombination radiation depends on the QMR. Hence, precise calculations of the radiative recombination that include non-standard processes are highly desired.

A quantitative analysis of the cosmological recombination requires a knowledge of wavefunctions that correctly describe an electron involved in the process in both the initial continuous and final discrete eigenstates. Especially challenging is to obtain the wavefunction in a closed algebraic form that correctly describes an electron in the initial continuous spectral state. To find this wavefunction, we employed an integral representation of the nonrelativistic Coulomb Green's function in parabolic coordinates. The wavefunctions of an electron in the bound state are also obtained in a closed algebraic form. The derived wavefunctions allow us to elaborate a convenient scheme of calculation, which can be used for the creation of a fast and complete cosmological code based on the quasi-molecular mechanism of recombination.

Preliminary research shows that the calculation of one free-bound transition probability takes a few seconds on a standard laptop. Results of calculations will be presented in forthcoming paper.

**Appendix A**

In parabolic coordinates the CGF can be represented as this expansion (Blinder 1981),

$$G^{(+)}(\vec{r},\vec{r}\,') = \frac{i}{2\pi^2 k} \sum_{m=-\infty}^{\infty} e^{im(\varphi-\varphi')} \int_{-\infty}^{\infty} d\lambda \, g_m^{(+)}(\beta-\lambda,\mu,\mu') g_m^{(+)}(\lambda,\nu,\nu'), \qquad (A1)$$

in which

$$g_m^{(+)}(\lambda,x,x') = \frac{\Gamma((m+1)/2 - i\lambda)}{\Gamma(m+1)} \frac{M_{i\lambda,m/2}(-ikx_<)}{(x_<)^{1/2}} \frac{W_{i\lambda,m/2}(-ikx_>)}{(x_>)^{1/2}}, \qquad (A2)$$

$M_{i\lambda,m/2}(-ikx_<)$ and $W_{i\lambda,m/2}(-ikx_>)$ represent Whittaker functions of the first and second kind, respectively, $\beta = Z_1/k$ and $x_< = \min(x,x')$, $x_> = \max(x,x')$.

Using the integral representation of the product of Whittaker functions (Gradshtein and Ryzhik 1980, Buchholz 1969), we obtain

$$g_m^{(+)}(\lambda,x,x') = (-1)^{\frac{m+1}{2}} k \int_0^\infty ds\, e^{i\frac{k}{2}(x+x')\cosh s} \left(\coth\frac{s}{2}\right)^{2i\lambda} J_m\left(k\sqrt{xx'}\sinh s\right), \tag{A3}$$

in which $J_m$ is a Bessel function.

Returning to equation (A1), integral representation (A3) can be applied to each factor $g_m^{(+)}(\beta-\lambda,\mu,\mu')g_m^{(+)}(\lambda,\nu,\nu')$. As a result we obtain that

$$G^{(+)}(\vec{r},\vec{r}') = -\frac{ik}{2\pi^2} \sum_{m=-\infty}^{\infty} (-1)^m e^{im(\varphi-\varphi')} \int_0^\infty ds\, e^{i\frac{k}{2}(\mu+\mu')\cosh s} J_m\left(k\sqrt{\mu\mu'}\sinh s\right)$$
$$\cdot \left(\coth\frac{s}{2}\right)^{2i\beta} \int_0^\infty dt\, e^{i\frac{k}{2}(\nu+\nu')\cosh t} J_m\left(k\sqrt{\nu\nu'}\sinh t\right) \int_{-\infty}^{\infty} d\lambda \left(\coth\frac{s}{2}\right)^{-2i\lambda} \left(\coth\frac{t}{2}\right)^{2i\lambda}. \tag{A4}$$

The integral over $\lambda$ gives a delta function,

$$G^{(+)}(\vec{r},\vec{r}') = -\frac{ik}{2\pi} \sum_{m=-\infty}^{\infty} (-1)^m e^{im(\varphi-\varphi')} \int_0^\infty ds\, e^{i\frac{k}{2}(\mu+\mu')\cosh s} J_m\left(k\sqrt{\mu\mu'}\sinh s\right)$$
$$\cdot \sinh s \left(\coth\frac{s}{2}\right)^{2i\beta} \int_0^\infty dt\, e^{i\frac{k}{2}(\nu+\nu')\cosh t} J_m\left(k\sqrt{\nu\nu'}\sinh t\right) \delta(t-s). \tag{A5}$$

Performing the integration over $t$ and taking into account that $(-1)^m J_m(x) = J_m(-x)$, we arrive at equation (4).

**Appendix B**

Here are defined functions that appear in equations (22):

$$A_{n_1,n_2,\pm m_f} = \frac{1}{\sqrt{2\pi}} \int_0^{2\pi} e^{\mp im_f\tilde{\varphi}} u^{\frac{m_f}{2}} F(-n_1,m_f+1,\gamma\alpha_1\tilde{\mu})F(-n_2,m_f+1,\gamma\alpha_2\tilde{\nu}) d\varphi,$$

$$B_{n_1,n_2,\pm m_f} = \frac{1}{\sqrt{2\pi}} \int_0^{2\pi} e^{\mp im_f\tilde{\varphi}} u^{\frac{m_f}{2}} Q(\tilde{\mu},\tilde{\nu}) F(-n_1,m_f+1,\gamma\alpha_1\tilde{\mu})F(-n_2,m_f+1,\gamma\alpha_2\tilde{\nu}) d\varphi,$$

$$C_{n_1,n_2,\pm m_f} = \frac{1}{\sqrt{2\pi}} \int_0^{2\pi} e^{\mp im_f\tilde{\varphi}} u^{\frac{m_f}{2}} F(-n_1,m_f+1,\gamma\alpha_1\tilde{\mu})F(-n_2,m_f+1,\gamma\alpha_2\tilde{\nu}) \cos\varphi\, d\varphi,$$

$$A_{n_1,n_2,\pm m_f}^{(\pm)} = \frac{1}{\sqrt{2\pi}} \int_0^{2\pi} e^{\pm i\varphi} e^{\mp im_f\tilde{\varphi}} u^{\frac{m_f}{2}} F(-n_1,m_f+1,\gamma\alpha_1\tilde{\mu})F(-n_2,m_f+1,\gamma\alpha_2\tilde{\nu}) d\varphi, \tag{B1}$$

$$B_{n_1,n_2,\pm m_f}^{(\pm)} = \frac{1}{\sqrt{2\pi}} \int_0^{2\pi} e^{\pm i\varphi} e^{\mp im_f\tilde{\varphi}} u^{\frac{m_f}{2}} Q(\tilde{\mu},\tilde{\nu}) F(-n_1,m_f+1,\gamma\alpha_1\tilde{\mu})F(-n_2,m_f+1,\gamma\alpha_2\tilde{\nu}) d\varphi.$$

$$C_{n_1,n_2,\pm m_f}^{(\pm)} = \frac{1}{\sqrt{2\pi}} \int_0^{2\pi} e^{\pm i\varphi} e^{\mp im_f\tilde{\varphi}} u^{\frac{m_f}{2}} F(-n_1,m_f+1,\gamma\alpha_1\tilde{\mu})F(-n_2,m_f+1,\gamma\alpha_2\tilde{\nu}) \cos\varphi\, d\varphi.$$

When $\vartheta_{\vec{R}} = 0$ these integrals are readily solvable

$$A_{n_1,n_2,\pm m_f}(0) = \sqrt{2\pi}(\mu\nu)^{\frac{m_f}{2}} F(-n_1,m_f+1,\gamma\alpha_1\mu)F(-n_2,m_f+1,\gamma\alpha_2\nu)\delta_{m_f,0},$$

$$B_{n_1,n_2,\pm m_f}(0) = \sqrt{2\pi}(\mu\nu)^{\frac{m_f}{2}} Q(\mu,\nu) F(-n_1,m_f+1,\gamma\alpha_1\mu)F(-n_2,m_f+1,\gamma\alpha_2\nu)\delta_{m_f,0},$$

$$C_{n_1,n_2,\pm m_f}(0) = \sqrt{\frac{\pi}{2}}(\mu\nu)^{\frac{m_f}{2}} Q(\mu,\nu) F(-n_1, m_f+1, \gamma\alpha_1\mu) F(-n_2, m_f+1, \gamma\alpha_2\nu)\delta_{m_f,1}, \quad (B2)$$

$$A^{(\pm)}_{n_1,n_2,\pm m_f}(0) = \sqrt{2\pi}(\mu\nu)^{\frac{m_f}{2}} F(-n_1, m_f+1, \gamma\alpha_1\mu) F(-n_2, m_f+1, \gamma\alpha_2\nu)\delta_{m_f,1},$$

$$B^{(\pm)}_{n_1,n_2,\pm m_f}(0) = \sqrt{2\pi}(\mu\nu)^{\frac{m_f}{2}} Q(\mu,\nu) F(-n_1, m_f+1, \gamma\alpha_1\mu) F(-n_2, m_f+1, \gamma\alpha_2\nu)\delta_{m_f,1},$$

$$C^{(\pm)}_{n_1,n_2,\pm m_f}(0) = \sqrt{\frac{\pi}{2}}(\mu\nu)^{\frac{m_f}{2}} Q(\mu,\nu) F(-n_1, m_f+1, \gamma\alpha_1\mu) F(-n_2, m_f+1, \gamma\alpha_2\nu)$$

$$\cdot(\delta_{m_f,0} + \delta_{m_f,2}).$$

When $\vartheta_{\tilde{R}} = \pi/2$ integrals can be substantially simplified in (C1) by making the appropriate transformations

$$A_{n_1,n_2,\pm m_f}(\pi/2) = \frac{1}{2^{m_f}\sqrt{2\pi}} \int_0^{2\pi} \left(\mu - \nu \pm 2i\sqrt{\mu\nu}\sin\varphi\right)^{m_f} F(-n_1, m_f+1, \gamma\alpha_1\tilde{\mu})$$
$$\cdot F(-n_2, m_f+1, \gamma\alpha_2\tilde{\nu})d\varphi,$$

$$B_{n_1,n_2,\pm m_f}(\pi/2) = \frac{1}{2^{m_f}\sqrt{2\pi}} \int_0^{2\pi} \left(\mu - \nu \pm 2i\sqrt{\mu\nu}\sin\varphi\right)^{m_f} F(-n_1, m_f+1, \gamma\alpha_1\tilde{\mu})$$
$$\cdot F(-n_2, m_f+1, \gamma\alpha_2\tilde{\nu})Q(\tilde{\mu},\tilde{\nu})d\varphi.$$

$$C_{n_1,n_2,\pm m_f}(\pi/2) = \frac{1}{2^{m_f}\sqrt{2\pi}} \int_0^{2\pi} \left(\mu - \nu \pm 2i\sqrt{\mu\nu}\sin\varphi\right)^{m_f} F(-n_1, m_f+1, \gamma\alpha_1\tilde{\mu}) \quad (B3)$$
$$\cdot F(-n_2, m_f+1, \gamma\alpha_2\tilde{\nu})\cos\varphi d\varphi,$$

$$A^{(\pm)}_{n_1,n_2,\pm m_f}(\pi/2) = \frac{1}{2^{m_f}\sqrt{2\pi}} \int_0^{2\pi} e^{\pm i\varphi}\left(\mu - \nu \pm 2i\sqrt{\mu\nu}\sin\varphi\right)^{m_f} F(-n_1, m_f+1, \gamma\alpha_1\tilde{\mu})$$
$$\cdot F(-n_2, m_f+1, \gamma\alpha_2\tilde{\nu})d\varphi,$$

$$B^{(\pm)}_{n_1,n_2,\pm m_f}(\pi/2) = \frac{1}{2^{m_f}\sqrt{2\pi}} \int_0^{2\pi} e^{\pm i\varphi}\left(\mu - \nu \pm 2i\sqrt{\mu\nu}\sin\varphi\right)^{m_f} F(-n_1, m_f+1, \gamma\alpha_1\tilde{\mu})$$
$$\cdot F(-n_2, m_f+1, \gamma\alpha_2\tilde{\nu})Q(\tilde{\mu},\tilde{\nu})d\varphi,$$

$$C^{(\pm)}_{n_1,n_2,\pm m_f}(\pi/2) = \frac{1}{2^{m_f}\sqrt{2\pi}} \int_0^{2\pi} e^{\pm i\varphi}\left(\mu - \nu \pm 2i\sqrt{\mu\nu}\sin\varphi\right)^{m_f} F(-n_1, m_f+1, \gamma\alpha_1\tilde{\mu})$$
$$\cdot F(-n_2, m_f+1, \gamma\alpha_2\tilde{\nu})\cos\varphi d\varphi.$$

For given quantum numbers $n_1$, $n_2$ and $m_f$ these integrals become a sum of analytically solvable integrals.

**Appendix C**

Integrals over $\mu$ that appear in equation (22) in terms proportional to $R^{-1}$ can be reduced to the integral

$$\upsilon_\tau^{(p)}(s,\mu') = \int_0^\infty e^{-\frac{\gamma - ik\cosh s}{2}\mu} J_\tau\left(k(\mu\mu')^{1/2}\sinh s\right)\mu^p d\mu, \quad (C1)$$

in which $\tau = 0$ or $\tau = 1$; $p = 0, 1, 2, \ldots$ for $\tau = 0$ and $p = 1/2, 3/2, 5/2, \ldots$ for $\tau = 1$. This integral is analytically solvable (Gradshtein and Ryzhik 1980)

$$\upsilon_\tau^{(p)}(s, \mu') = \frac{2^{p+1-\tau/2} \Gamma(p+1+\tau/2)(k \sinh s)^\tau \mu'^{\tau/2}}{\Gamma(\tau+1)(\gamma - ik \cosh s)^{p+1+\tau/2}}$$
$$\cdot {}_1F_1\left(p+1+\frac{\tau}{2}, \tau+1, -\frac{(k \sinh s)^2 \mu'}{2(\gamma - ik \cosh s)}\right). \quad (C2)$$

In the confluent hypergeometric function the first parameter is equal to the second one or is greater by an integer, $p + 1 + \tau/2 = \tau + 1 + l$ where $l = 0, 1, 2, \ldots$. Taking into account that ${}_1F_1(\tau+1, \tau+1, x) = e^x$ and making use a recurrence relation (Janke, Emde, Lösch, 1960)

$${}_1F_1(a+1, \tau+1, x) = a^{-1}\left[(2a - \tau - 1 + x) {}_1F_1(a, \tau+1, x) + (\tau + 1 - a) {}_1F_1(a-1, \tau+1, x)\right], \quad (C3)$$

${}_1F_1(\tau+1+l, \tau+1, x)$ can be represented as a product of $e^x$ and a polynomial function of order $l$.

Inserting $\upsilon_\tau^{(p)}(s, \mu')$ into integrals over $\mu'$, we arrive at a sum of analytically solvable integrals of this type

$$\chi_\tau^{(p')}(s) = \int_0^{R/2} e^{-\frac{\gamma}{2}\left(\frac{(k \sinh s)^2}{\gamma^2 + (k \cosh s)^2}\right)\mu'} \left[\cos(\phi\mu') + i \sin(\phi\mu')\right] \mu'^{(p'+\tau/2)} d\mu', \quad (C4)$$

in which $p' + \tau/2 = 0, 1, 2, \ldots$ and $\phi = 2^{-1} k \left[1 + \cosh s \left(1 - (k \sinh s)^2 / (\gamma^2 + (k \cosh s)^2)\right)\right]$.

Integrals over $\nu$ that appear in equation (22) in terms proportional to $R^{-1}$

$$h_\tau^{(\rho)}(s, \nu') = \int_0^\infty e^{-\frac{\gamma - ik \cosh s}{2}\nu} J_\tau\left(-k(\nu\nu')^{1/2} \sinh s\right) \nu^\rho d\nu, \quad (C5)$$

in which $\rho = 0, 1, 2, \ldots$ for $\tau = 0$ and $\rho = 1/2, 3/2, 5/2, \ldots$ for $\tau = 1$ can be solved as (C1). Indeed, taking into account that $J_\tau(-x) = (-1)^\tau J_\tau(x)$, we obtain

$$h_\tau^{(\rho)}(s, \nu') = (-1)^\tau \frac{2^{\rho+1-\tau/2} \Gamma(\rho+1+\tau/2)(k \sinh s)^\tau \nu'^{\tau/2}}{\Gamma(\tau+1)(\gamma - ik \cosh s)^{\rho+1+\tau/2}}$$
$$\cdot {}_1F_1\left(\rho+1+\frac{\tau}{2}, \tau+1, -\frac{(k \sinh s)^2 \nu'}{2(\gamma - ik \cosh s)}\right). \quad (C6)$$

In (C6) the confluent hypergeometric function can be represented as a product of an exponential and a polynomial functions.

Inserting $h_\tau^{(\rho)}(s, \nu')$ into integrals over $\nu'$, we come to a sum of integrals of this type

$$g_\tau^{(\rho')}(s) = \int_0^{R/2} e^{-\frac{1}{2}\left[\frac{(k \sinh s)^2}{\gamma - ik \cosh s} + ik(1 - \cosh s)\right]\nu'} {}_1F_1\left(\frac{i}{k}, 1, ik\nu'\right) \nu'^{(\rho'+\tau/2)} d\nu', \quad (C7)$$

in which $\rho' + \tau/2 = 0, 1, 2, \ldots$. These integrals, we calculate numerically.